\RequirePackage{fix-cm}
\documentclass[smallextended]{svjour3}       % onecolumn (second format)
\smartqed  % flush right qed marks, e.g. at end of proof

\usepackage{graphicx}
\usepackage{subfigure}
\usepackage{amssymb}
\usepackage{color}
\usepackage{amsmath}
\usepackage{cite}
\usepackage{graphicx}
\usepackage{epstopdf}
\usepackage{cases}
 \journalname{Nonlinear Dynamics}
\begin{document}

\title{Reductions of the $(4+1)$-dimensional Fokas equation and their solutions
\thanks{$^{*}$Corresponding author, Jingsong He: hejingsong@szu.edu.cn; jshe@ustc.edu.cn}
}
\subtitle{}

%\titlerunning{Short form of title}        % if too long for running head

\author{Yulei Cao \and  Jingsong He$^{*}$ \and Yi Cheng \and Dumitru Mihalache
}

%\authorrunning{Short form of author list} % if too long for running head

\institute{Yulei Cao, Yi Cheng \at
              School of Mathematical Sciences, USTC, Hefei, Anhui 230026, P.\ R.\ China
\and Jingsong He \at
              Institute for Advanced Study, Shenzhen University, Shenzhen, Guangdong 518060, P.\ R.\ China
              %\email{hejingsong@szu.edu.cn; jshe@ustc.edu.cn}
\and Dumitru Mihalache \at Horia Hulubei National Institute of Physics and Nuclear Engineering, Magurele, RO 077125,
Romania
}

\date{Received: date / Accepted: date}
% The correct dates will be entered by the editor

\maketitle

\begin{abstract}
An integrable extension of the Kadomtsev-Petviashvili (KP)
and Davey-Stewartson (DS) equations is investigated in this paper.
We will refer to this integrable extension as the $(4+1)$-dimensional
Fokas equation. The determinant expressions of soliton,
breather, rational, and semi-rational solutions
of the $(4 + 1)$-dimensional Fokas equation are constructed
based on the Hirota's bilinear method and the KP hierarchy
reduction method.  The complex dynamics of these new exact solutions
are shown in both three-dimensional plots and two-dimensional
contour plots. Interestingly, the
patterns of obtained high-order lumps are similar to
those of rogue waves in the $(1+1)$-dimensions by choosing
different values of the free parameters of the model.
Furthermore, three kinds of new semi-rational solutions are
presented and
the classification of lump fission and fusion processes is
also discussed. Additionally,
we give a new way to obtain rational and semi-rational
solutions of $(3+1)$-dimensional KP equation
by reducing the solutions of the $(4+1)$-dimensional
Fokas equation. All these results show that the
$(4+1)$-dimensional Fokas equation is a meaningful
multidimensional extension of the KP and DS equations.
The obtained results might be useful in diverse fields such as
hydrodynamics, nonlinear optics and photonics,
ion-acoustic waves in plasmas, matter waves in Bose-Einstein condensates,
and sound waves in ferromagnetic media.

\keywords{ 4D Fokas equation $\cdot$ KP hierarchy reduction method
$\cdot$ Rational solution $\cdot$ Semi-rational solution
$\cdot$ 3D KP equation
}
% \PACS{PACS code1 \and PACS code2 \and more}
% \subclass{MSC code1 \and MSC code2 \and more}
\end{abstract}

\section{Introduction}

The research field of solitary waves is in fact
an interdisciplinary research area that has been deeply studied
both theoretically and experimentally. Solitary waves in hydrodynamics originated
from the accidental discovery of Russell in 1834 \cite{js1}.
Nevertheless, he failed to give a rigorous proof
of the existence of such special type of waves. More than
60 years later, Korteweg and de Vries \cite{kdv} made
a comprehensive analysis of these solitary waves and
established a mathematical model of shallow water
waves that adequately describes their complex dynamics.  About 70 years later, in 1965, Zabusky and Kruskal \cite{nj} found numerically that
such solitary waves have the property of
elastic scattering,
and called them "solitons". This pioneering work of Zabusky and Kruskal was a
milestone in the area of solitons and all that. Since then, the study
of solitons has begun to flourish in various fields
such as nonlinear optics and optical
fibers\cite{liu1,liu2}, condensed matter, fluid mechanics, and
plasma physics. The current research mainly focuses
on $(1+1)$-dimensional (1D) and
$(2+1)$-dimensional (2D) systems.
However the physical space in reality is
$(3+1)$-dimensional (3D) and one of the most important
open problems in soliton theory is to construct
integrable nonlinear partial differential equations (NPDEs) in higher than
two spatial dimensions. Therefore, the research value
of high-dimensional nonlinear systems is enormous.
In order to seek new high-dimensional
integrable NPDEs,
many researchers have made great efforts during the past decades \cite{ab1,
ds1, jm, lou1, lou2, ytsf, geng, fokas1, fokas2, fokas3, fokas5, fokas6,
fokas7, ND1, ND2, ND3, ND4, ND5, ChenMihalache, KaurWazwaz, MM}.
But, there are still many meaningful open problems
to be addressed. With the increasing number of
variables, solving high-dimensional NPDEs will be
very difficult. Therefore, it is a challenging work
to obtain exact solutions of high-dimensional
nonlinear systems. Furthermore, a natural problem
is whether the exact solutions of  high-dimensional NPDEs
can be reduced to the exact solutions of
low-dimensional NPDEs?

Inspired by the above problems, we consider
the $(4 + 1)$-dimensional (4D) Fokas equation\cite{fokas}:
\begin{equation}\label{N1}
\begin{aligned}
u_{x_1t}-\frac{1}{4}u_{x_1x_1x_1x_2}+\frac{1}{4}u_{x_1x_2x_2x_2}
+\frac{3}{2}(u^2)_{x_1x_2}-\frac{3}{2}u_{y_1y_2}=0.
\end{aligned}
\end{equation}
This equation was introduced by Fokas in 2006 \cite{fokas},
being an integrable extension of the Kadomtsev-Petviashvili (KP)
and Davey-Stewartson (DS) equations. Because of
the important physical applications of KP and
DS equations, the 4D Fokas equation may be used to
describe surface and internal waves in rivers
with different physical situations. Solitons \cite{fs1,fs2},
quasi-periodic solutions \cite{JMP}, lumps \cite{fl1,fl2}
and lump-soliton solutions \cite{ff1} for the 4D Fokas equation
have been investigated. However, these studies
are far from being complete.  To the best of
authors' knowledge, high-order rational and
semi-rational solutions for the 4D Fokas equation
have never been reported. In this paper, we mainly
focus on the new exact solutions of the 4D Fokas equation,
and how to reduce the exact solutions of the 4D Fokas equation
to the exact solutions of low-dimensional NPDEs.

The structure of this paper is as follows.
In section \ref{2}, the determinant expressions
of soliton and breather solutions are constructed
by using the KP hierarchy reduction method.
In sections \ref{3} and \ref{4}, high-order
rational and semi-rational solutions are generated
for  the 4D Fokas equation and the complex dynamic behavior
of the corresponding solutions are shown by
both three-dimensional plots and two-dimensional contour plots.
Then in sections \ref{5},
a new way for obtaining rational and
semi-rational solutions of 3D KP equation
is presented. Finally, in section \ref{7} we
discuss and summarize our results.

\section{Soliton and breather solutions in the determinant form} \label{2}

In this section, we introduce the determinant
expression of soliton and breather solutions
for the 4D Fokas equation. Through the
following transformation:
\begin{equation*}
\begin{aligned}
x=k_1x_1+k_2x_2,
\end{aligned}
\end{equation*}
the 4D Fokas equation \eqref{N1} becomes the following 3D equation
\begin{equation}
\begin{aligned}
u_{xt}+\frac{1}{4}k_2(k^2_2-k^2_1)u_{xxxx}+\frac{3k_2}{2}(u^2)_{xx}
-\frac{3}{2k_1}u_{y_1y_2}=0.
\end{aligned}
\end{equation}
%%%%%%%%%%%%%%%%%%%%%%%%%
%%%%%%%%%%%%%%%%%%%%%%%%%
Additionally, if we further make the transformation
\begin{equation*}
\begin{aligned}
y=k_3y_1+k_4y_2,
\end{aligned}
\end{equation*}
then the 4D Fokas equation becomes
\begin{equation}\label{N2}
\begin{aligned}
u_{xt}+\frac{1}{4}k_2(k^2_2-k^2_1)u_{xxxx}+\frac{3k_2}{2}(u^2)_{xx}
-\frac{3k_3k_4}{2k_1}u_{yy}=0.
\end{aligned}
\end{equation}

Now we make the variable transformation
\begin{equation}
\begin{aligned}
u=(k^2_2-k^2_1)(\ln f)_{xx}.
\end{aligned}
\end{equation}
Then the 4D Fokas equation \eqref{N1} is transformed
into the following bilinear form:
\begin{equation}\label{bilinear}
\begin{aligned}
&[D^{4}_{x}+\frac{4}{k_2}(k^2_2-k^2_1)D_{x}D_{t}
-\frac{6k_3k_4}{k_1k_2}(k^2_2-k^2_1)D^2_{y}]f \cdot f =0,\\
\end{aligned}
\end{equation}
where $D$ is Hirota's bilinear differential operator \cite{hirota}.
Applying the change of independent variables
\begin{equation}\label{change}
\begin{aligned}
 z_{1}=x,\quad z_{2}=\sqrt{\frac{{k_1k_2}(k^2_2-k^2_1)}{2k_3k_4}}iy,
 \quad z_{3}=-2k_2(k^2_2-k^2_1)t,
\end{aligned}
\end{equation}
the bilinear form \eqref{bilinear} can be transformed
into the following bilinear equation of the
KP hierarchy \cite{jm}:
\begin{equation}\label{kphierarchy}
\begin{aligned}
&[D^{4}_{z_1}-4D_{z_{1}}D_{z_{3}}+3D^2_{z_{2}}]f \cdot f =0.
\end{aligned}
\end{equation}
According to Sato theory \cite{yo,yo1}, we construct
the Gram determinant solutions of the 4D Fokas equation.

\textbf{Theorem 1.}  The 4D Fokas equation \eqref{N1}
admits the following soliton and breather solutions:
\begin{equation}\label{soliton}
\begin{aligned}
u=(k^2_2-k^2_1)(\ln f)_{xx},\quad x=k_1x_1+k_2x_2,
\end{aligned}
\end{equation}
with
\begin{equation}
\begin{aligned}
&f=\det_{1\leq i,j\leq N}(m^{(0)}_{i,j}), \quad  {\rm where} \quad m^{(n)}_{i,j}=\delta_{ij}+\frac{1}{p_{i}+p^*_{j}}(-\frac{p_i}{p^*_j})^{n} e^{\xi_i +\xi^*_j},\\
\end{aligned}
\end{equation}
\begin{equation*}
\begin{aligned}
\xi_j&=k_1p_ix_1+k_2p_ix_2+\sqrt{\frac{{k_1k_2}
(k^2_2-k^2_1)}{-2k_3k_4}}p_i^2y_1\\
&+\sqrt{\frac{{k_1k_2}
(k^2_2-k^2_1)}{-2k_3k_4}}p_i^2y_2-k_2(k^2_2-k^2_1)p_i^3t+\xi_{i0}.
\end{aligned}
\end{equation*}
Here $\delta_{ij}=0,1$, $p_i$ and $\xi_{i0}$ are
arbitrary complex constants, $i, j$, and $N$ are
arbitrary positive integers, and the asterisk denotes
the complex conjugation. We must emphasize that
$\frac{{k_1k_2}(k^2_2-k^2_1)}{-2k_3k_4}>0$ must hold.
In order to prove Theorem 1, we first introduce the following Lemma.

\textbf{Lemma 1.}  The bilinear equation of KP
hierarchy \eqref{kphierarchy} has solutions
\begin{equation}\label{B2s}
\begin{aligned}
\tau_{n}=\det_{1\leq i,j\leq N}(m^{(n)}_{i,j}),\\
\end{aligned}
\end{equation}
with the matrix element $m^{(n)}_{i,j}$ satisfying
the following differential and difference relations
\begin{equation}\label{lemma1}
\begin{aligned}
&\partial_{z_{1}}m^{(n)}_{i,j}=\varphi^{(n)}_{i}\psi^{(n)}_{j},\\
&m^{(n+1)}_{i,j}=m^{(n)}_{i,j}+\varphi^{(n)}_{i}\psi^{(n+1)}_{j},\\
&\partial_{z_{2}}m^{(n)}_{i,j}=\varphi^{(n+1)}_{i}\psi^{(n)}_{j}
+\varphi^{(n)}_{i}\psi^{(n-1)}_{j},\\
&\partial_{z_{3}}m^{(n)}_{i,j}=\varphi^{(n+2)}_{i}\psi^{(n)}_{j}
+\varphi^{(n+1)}_{i}\psi^{(n-1)}_{j}
+\varphi^{(n)}_{i}\psi^{(n-2)}_{j},\\
&\partial_{z_{k}}\varphi=\varphi^{(n+k)}_{i}, \quad \partial_{z_{k}}\psi_{i}
=-\psi^{(n-k)}_{i},(k=1,2,3).\\
\end{aligned}
\end{equation}
Here  $m^{(n)}_{i,j}$,$\varphi^{(n)}_{i}$, and $\psi^{(n)}_{j}$
are functions of the variables $z_{1},z_{2}$, and $z_{3}$.

\textbf{Proof of Lemma 1.}
Reusing the differential of determinant and
the expansion formula of bordered determinant \cite{yo,yo1},
the derivatives of the $\tau$ functions
can be expressed by the following bordered
determinants:
\begin{align*}
& \partial_{z_{1}}\tau_{n}=\begin{vmatrix} m^{(n)}_{i,j} & \varphi^{(n)}_{i}
 \\ -\psi^{(n)}_{j} & 0 \end{vmatrix} ,\\
& \partial_{z_{2}}\tau_{n}=\begin{vmatrix}m^{(n)}_{i,j} & \varphi^{(n+1)}_{i}
\\ -\psi^{(n)}_{j} & 0\end{vmatrix} +
\begin{vmatrix} m^{(n)}_{i,j} & \varphi^{(n)}_{i} \\ -\psi^{(n-1)}_{j} & 0 \end{vmatrix},\\
& \partial_{z_{3}}\tau_{n}=\begin{vmatrix} m^{(n)}_{i,j} & \varphi^{(n+2)}_{i}
\\ -\psi^{(n)}_{j} & 0 \end{vmatrix} + \begin{vmatrix} m^{(n)}_{i,j} & \varphi^{(n+1)}_{i}
\\ -\psi^{(n-1)}_{j} & 0\end{vmatrix}+\begin{vmatrix} m^{(n)}_{i,j} & \varphi^{(n)}_{i}
 \\ -\psi^{(n-2)}_{j} & 0 \end{vmatrix},\\
&\partial_{z_{1}}\partial_{z_{3}}\tau_{n}=\begin{vmatrix}m^{(n)}_{i,j} & \varphi^{(n+3)}_{i}
\\ -\psi^{(n)}_{j} & 0\end{vmatrix} +\begin{vmatrix} m^{(n)}_{i,j} & \varphi^{(n)}_{i}
\\ \psi^{(n-3)}_{j} & 0 \end{vmatrix},\\
& \partial^2_{z_{2}}\tau_{n}=\begin{vmatrix}m^{(n)}_{i,j} & \varphi^{(n+3)}_{i}
 \\ -\psi^{(n)}_{j} & 0\end{vmatrix} + \begin{vmatrix} m^{(n)}_{i,j} & \varphi^{(n+1)}_{i}
 \\ \psi^{(n-2)}_{j} & 0 \end{vmatrix}+\begin{vmatrix}m^{(n)}_{i,j} & \varphi^{(n+2)}_{i}
 \\ -\psi^{(n-1)}_{j} & 0\end{vmatrix}+\begin{vmatrix}m^{(n)}_{i,j} & \varphi^{(n)}_{i}
 \\ \psi^{(n-3)}_{j} & 0\end{vmatrix},\\
& \partial^4_{z_{1}}\tau_{n}=\begin{vmatrix}m^{(n)}_{i,j} & \varphi^{(n+3)}_{i}
\\ -\psi^{(n)}_{j} & 0\end{vmatrix} + 3\begin{vmatrix} m^{(n)}_{i,j} & \varphi^{(n+2)}_{i}
\\ \psi^{(n-1)}_{j} & 0 \end{vmatrix}+3\begin{vmatrix}m^{(n)}_{i,j} & \varphi^{(n+1)}_{i}
\\ -\psi^{(n-2)}_{j} & 0\end{vmatrix}+\begin{vmatrix}m^{(n)}_{i,j} & \varphi^{(n)}_{i}
\\ \psi^{(n-3)}_{j} & 0\end{vmatrix}.\\
\end{align*}
As a result:
$$(\partial^3_{z_{1}}-4\partial_{z_{1}}\partial_{z_{3}}
+3\partial^2_{z_{2}})\tau_{n} \times \tau_{n}=0,$$
$$4\partial_{z_{1}}\tau_{n} \times (\partial_{z_{3}}\tau_{n}
-\partial^3_{z_{1}}\tau_{n})+6(\partial^2_{z_{1}}\tau_{n})^2
- 6(\partial_{z_{2}}\tau_{n})^2=0.$$
This completes the proof of Lemma 1.
Then, we will prove Theorem 1 with Lemma 1.

\textbf{Proof of Theorem 1.} In order to construct
soliton and breather solutions
for the bilinear equation \eqref{kphierarchy},
we choose functions
$m^{(n)}_{i,j}$, $\varphi^{(n)}_{i}$,
and $\psi^{(n)}_{j}$  as follows
\begin{equation}
\begin{aligned}
&m^{(n)}_{i,j}=\delta_{ij}+\frac{1}{p_{i}+q_{j}}\varphi^{(n)}_{i}\psi^{(n)}_{j},\\
&\varphi^{(n)}_{i}=p^{n}_{i}e^{\xi^{i}},\\
& \psi^{(n)}_{j}=(-q_{j})^{-n}e^{\eta_{j}},\\
\end{aligned}
\end{equation}
where
\begin{equation*}
\begin{aligned}
\xi_i=p_iz_1+p_i^2z_2+p_i^3z_3+\xi_{i0},\\
\eta_j=q_jz_1-q_j^2z_2+q_j^3z_3+\eta_{j0},
\end{aligned}
\end{equation*}
and $p_i, q_j, \xi_{i0}$, and $\eta_{j0}$ are
arbitrary complex constants.
Through the following restrictions:$$z_{1}=k_1x_1+k_2x_2,
\quad z_{2}=\sqrt{\frac{{k_1k_2}
(k^2_2-k^2_1)}{2k_3k_4}}i(k_3y_1+k_4y_2),$$
$$z_{3}=-2k_2(k^2_2-k^2_1)t,\quad  q^*_j=p_j,
\quad \eta^*_{j0}=\xi_{j0},$$
and then setting $f=\tau_{0}$, $\delta_{ij}=0,1$,
the solutions of bilinear equation \eqref{kphierarchy}
can be transformed into the solutions of the 4D Fokas equation.
This completes the proof of Theorem 1. Without losing
generality, we take $k_1=1, k_2=2, k_3=1$, and $k_4=1$
in this section.

\subsection{$N$-soliton solutions}

\begin{figure}[!htb]
\centering
\subfigure[t=0]{\includegraphics[height=3cm,width=3.8cm]{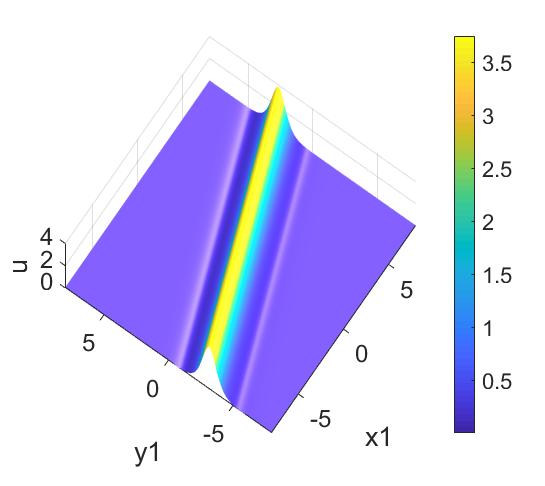}}
\subfigure[t=0]{\includegraphics[height=3cm,width=3.8cm]{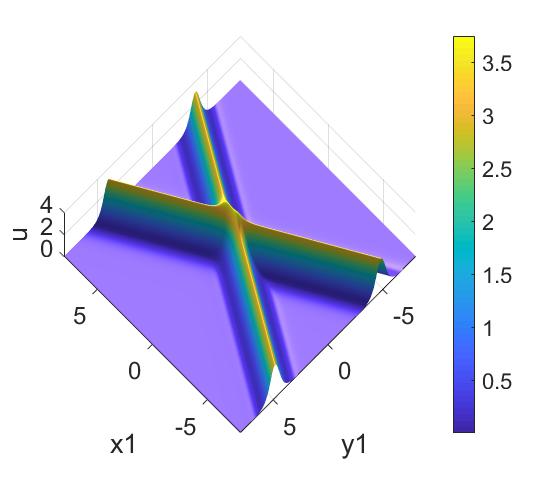}}
\subfigure[t=0]{\includegraphics[height=3cm,width=3.8cm]{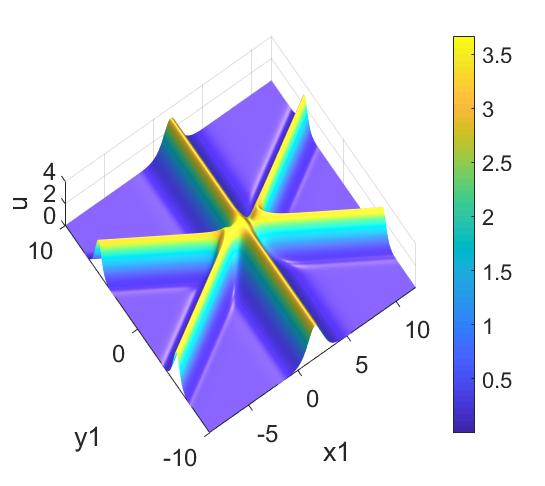}}
\caption{Dynamic behavior of solutions of the 4D Fokas
equation defined in equation \eqref{soliton}. (a): one-soliton solution with parameters
$N=1, \delta_{11}=1, p_1=1-\frac{i}{2}, x_2=0, y_2=0$, and $t=0$; (b): two-soliton solution with parameters
$N=2, \delta_{11}=1, \delta_{12}=0,\delta_{21}=0,
\delta_{22}=1, p_1=1-\frac{i}{2}, p_2=1+\frac{i}{2},
x_2=0, y_2=0$, and $t=0$; (c): three-soliton solution with parameters
$N=3, \delta_{jj}=1, \delta_{ij}=0 (i,j=1,2,3),
p_1=1-\frac{i}{2}, p_2=1+\frac{i}{2}, p_3=1,
x_2=0, y_2=0$, and $t=0$.}
\label{f1}
\end{figure}

Equation \eqref{N1} admits $N$-soliton solutions, assuming
 $\delta_{ij}=1$ when $i=j$, and $\delta_{ij}=0$ when
 $i\neq j$ in \eqref{soliton}. The one-soliton solution $u_{\rm 1s}$ is
generated by taking $N=1$ and $p_1=p_{1R}-ip_{1I}$:
\begin{equation}\label{1s}
\begin{aligned}
u_{\rm 1s}=\frac{24p_{1R}^3e^{2p_{1R}}[ x_1+2x_2-2\sqrt{3}p_{1I}(y_1
+y_2)-6(p_{1R}^2-3p_{1I}^2)t]}
{\left (2p_{1R}+e^{2p_{1R}}[ x_1+2x_2-2\sqrt{3}p_{1I}(y_1
+y_2)-6(p_{1R}^2-3p_{1I}^2)t]\right )^2}.
\end{aligned}
\end{equation}
From the above expressions, it is not difficult to calculate
that the maximum amplitude of the one-soliton solution is $3p^2_{1R}$;
when $x_1,y_1$ $\longrightarrow$ $\pm \infty$ solution
$u_{\rm 1s}$ approaches to the constant background plane
$0$ in the $(x_1,y_1)$-plane [see Fig. \ref{f1}(a)].
The velocity and center of the soliton are $6p_{1R}^2-18p_{1I}^2$
and $x_1+2x_2-2\sqrt{3}p_{1I}(y_1+y_2)-6(p_{1R}^2-3p_{1I}^2)t$,
respectively. By taking the parameters $N=2, p_1=1-\frac{i}{2}$
and $p_2=1+\frac{i}{2}$ in equation \eqref{soliton}
we obtain the expression of the two-soliton solution
$u_{\rm 2s}$ [see Fig. \ref{f1}(b)]:
\begin{equation}\label{2s}
\begin{aligned}
u_{\rm 2s}=\frac{120e^{\iota_{11}}+120e^{\iota_{12}}+960e^{\iota_{13}}
+2400e^{\iota_{14}}+2400e^{\iota_{15}}}
{\left (e^{\zeta_{11}}+10e^{\zeta_{12}}+10e^{\zeta_{13}}+20e^{\zeta_{14}}\right)^2},
\end{aligned}
\end{equation}
%%%%%%%%%%%%%%%%
\begin{equation*}
\begin{aligned}
\zeta_{11}&=4x_1+8x_2+2\sqrt{3}y_1+2\sqrt{3}y_2, \quad \zeta_{14}=2\sqrt{3}y_1+2\sqrt{3}y_2+6t,\\
\zeta_{12}&=2x_1+4x_2+4\sqrt{3}y_1+4\sqrt{3}y_2+3t, \quad \zeta_{13}=2x_1+4x_2+3t,\\
\iota_{11}&=\zeta_{11}+\zeta_{12},\qquad \iota_{12}=\zeta_{11}+\zeta_{13}, \qquad
\iota_{13}=\zeta_{12}+\zeta_{13}, \\
\iota_{14}&=\zeta_{12}+\zeta_{14}, \quad \quad \iota_{15}=\zeta_{13}+\zeta_{14}.
\end{aligned}
\end{equation*}

Additionally, taking the parameters $N=3, p_1=1-\frac{i}{2}$
, $p_2=1+\frac{i}{2}$, and $p_3=1$ in equation \eqref{soliton},
the three-soliton solution is obtained.
We also give the expression of the three-soliton solution in
the $(x_1,y_1)$-plane [see Fig. \ref{f1}(c)] in which $f$
is expressed as
\begin{equation}\label{3s}
\begin{aligned}
f=&1+\frac{1}{2}e^{2x_1+4x_2+2\sqrt{3}y_1+2\sqrt{3}y_2-3t}+\frac{1}{68}e^{4x_1+8x_2
+2\sqrt{3}y_1+2\sqrt{3}y_2-15t}\\
&+\frac{1}{68}e^{4x_1+8x_2-2\sqrt{3}y_1-2\sqrt{3}y_2-15t}
+\frac{1}{2}e^{2x_1+4x_2-2\sqrt{3}y_1-2\sqrt{3}y_2-3t}\\
&+\frac{1}{11560}e^{6x_1+12x_2-18t}+\frac{1}{2}e^{2x_1+4x_2-12t}+\frac{1}{20}e^{4x_1+8x_2-6t}.
\end{aligned}
\end{equation}

\subsection{A hybrid of a V-type soliton and breathers}

\begin{figure}[!htb]
\centering
\subfigure[t=0]{\includegraphics[height=3cm,width=4.5cm]{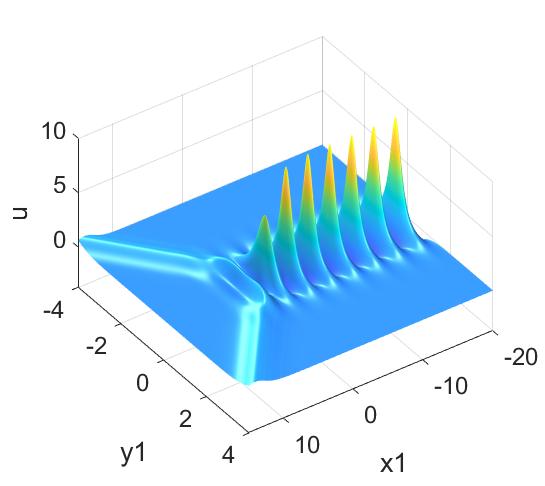}}\qquad
\subfigure[t=0]{\includegraphics[height=3cm,width=4.5cm]{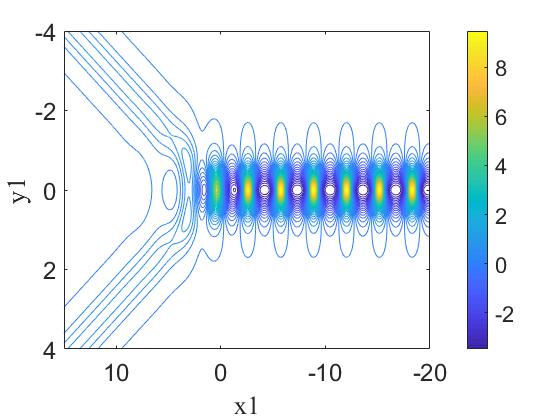}}
\caption{A hybrid of a V-type soliton and one breather of the 4D Fokas
equation with parameters $N=2, p_1=\frac{1}{2}+i, p_2=\frac{1}{2}-i,
\delta_{ij}=1 \, (i,j=1,2), x_2=0, y_2=0$, and $t=0$ in \eqref{soliton}. Panel (b)
is the contour plot of panel (a). }
\label{f2}
\end{figure}
In addition to the soliton solutions, equation \eqref{N1} admits
a hybrid of V-type soliton and
breather solutions, assuming $N\geq 2$ and $\delta_{ij}=1$ and
some parameters $p_i$ are complex in equation \eqref{soliton}.
We first consider the case of $N=2$ and $\delta_{ij}=1$. The following
parameters are further taken in equation \eqref{soliton}:
$$p_1=\frac{1}{2}+i, \quad p_2=\frac{1}{2}-i, \quad \delta_{12}=1,\quad \delta_{21}=1,$$
and the mixed solution consisting of a V-type soliton and one
breather solution is derived, see Fig. \ref{f2}. For this mixed solution the expression of $f$ is as follows:
\begin{equation}
\begin{aligned}
f=&\cosh[\gamma_1-2\sqrt{3}(y_1+y_2)]+\sinh[\gamma_1-2\sqrt{3}(y_1+y_2)]
+\cosh[\gamma_1+2\sqrt{3}(y_1+y_2)]\\
&+\sinh[\gamma_1-2\sqrt{3}(y_1+y_2)]+\frac{4}{5}[\cosh(2\gamma_1)+\sinh(2\gamma_1)]
-\frac{4}{5}[\cosh(\gamma_1)\\
&+\sinh(\gamma_1)]\sin(\gamma_2)-\frac{2}{5}[\cosh(\gamma_1)
+\sinh(\gamma_1)]\cos(\gamma_2),
\end{aligned}
\end{equation}
where
\begin{equation}
\begin{aligned}
\gamma_1=x_1+2x_2+\frac{33}{2}t,\quad \gamma_2=2x_1+4x_2+3t.
\end{aligned}
\end{equation}

\begin{figure}[!htb]
\centering
\subfigure[t=0]{\includegraphics[height=3cm,width=5cm]{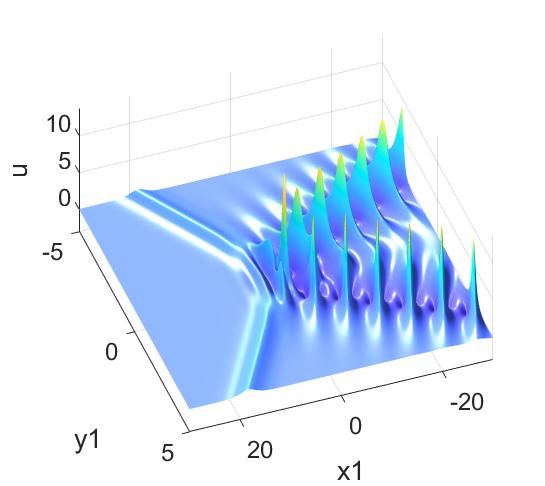}}\quad
\subfigure[t=0]{\includegraphics[height=3cm,width=5cm]{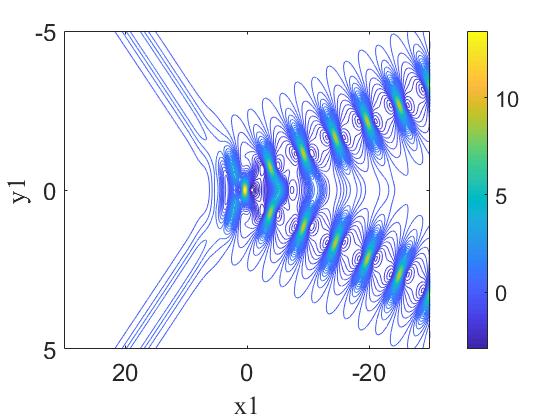}}
\caption{A hybrid of a V-type soliton and two breathers of the 4D Fokas
equation with parameters $N=3, p_1=\frac{1}{2}+i, p_2=\frac{1}{2}-i, p_3=\frac{2}{3},
\delta_{ij}=1 \,(i,j=1,2,3), x_2=0, y_2=0$, and $t=0$ in \eqref{soliton}. Panel (b)
is the contour plot of panel (a). }
\label{f2b}
\end{figure}

Furthermore, for larger $N$, we can derive the mixed solution
consisting of a V-type soliton and more breathers.
For example, when we take
the parameters $N=3, p_1=\frac{1}{2}+i,p_2=\frac{1}{2}-i,p_3=\frac{2}{3}$
and $\delta_{ij}=1 \, (i,j=1,2,3)$ in equation \eqref{soliton}
the mixed solution consisting of
a V-type soliton and two breathers is presented in  Fig. \ref{f2b}.

\section{Rational solutions in the determinant form} \label{3}

The rational solutions of
low-dimensional integrable systems have been
extensively investigated. However, there are
few studies of rational solutions in  high-dimensional systems.
Inspired by the works of
Ohta and Yang \cite{yo1,yo2,yo3}, the rational solutions of the
$\rm 4D$ Fokas equation are constructed by
introducing the following Lemma.

\textbf{Lemma 2.}  The bilinear equation of KP hierarchy \eqref{kphierarchy} has solutions
\begin{equation}
\begin{aligned}
\tau^{'}_{n}=\det_{1\leq i,j\leq N}(M^{(n)}_{2i-1,2j-1}),\\
\end{aligned}
\end{equation}
with the matrix element $M^{(n)}_{i,j}$ satisfying the following
differential and difference relations
\begin{equation}
\begin{aligned}
&\partial_{z_{1}}M^{(n)}_{i,j}=\Phi^{(n)}_{i}\Psi^{(n)}_{j},\\
&m^{(n+1)}_{i,j}=M^{(n)}_{i,j}+\Phi^{(n)}_{i}\Psi^{(n+1)}_{j},\\
&\partial_{z_{2}}M^{(n)}_{i,j}=\Phi^{(n+1)}_{i}\Psi^{(n)}_{j}
+\Phi^{(n)}_{i}\Psi^{(n-1)}_{j},\\
&\partial_{z_{3}}M^{(n)}_{i,j}=\Phi^{(n+2)}_{i}\Psi^{(n)}_{j}
+\Phi^{(n+1)}_{i}\Psi^{(n-1)}_{j}
+\Phi^{(n)}_{i}\Psi^{(n-2)}_{j},\\
&\partial_{z_{k}}\Phi=\Phi^{(n+k)}_{i},\quad \partial_{z_{k}}\Psi_{i}
=-\Psi^{(n-k)}_{i},(k=1,2,3).\\
\end{aligned}
\end{equation}
Here $M^{(n)}_{i,j}$, $\Phi^{(n)}_{i}$, and $\Psi^{(n)}_{j}$
are functions of the variables $z_{1}, z_{2}$, and $z_{3}$. The above relations
were proven in \cite{yo1}, hence we omit here the proof.
The functions $\Phi_i^{(n)}$, $\Psi_j^{(n)}$, and $M_{ij}^{(n)}$ are defined by
\begin{equation}\label{r1}
\begin{aligned}
\Phi_i^{(n)}&=A_ip^ne^{\xi}, \quad \Psi_j^{(n)}=B_j(-q)e^{\eta},\\
M^{(n)}_{i,j}&= \int^{z_1}\Phi_i^{(n)}\Psi_j^{(n)}dz_1=
A_iB_j\frac{1}{p+q}(-\frac{p}{q})^ne^{\xi+\eta},\\
\end{aligned}
\end{equation}
where
\begin{equation}
\begin{aligned}
A_i=&\sum_{k=0}^{i}\frac{c_{k}}{(i-k)!}(p\partial_{p})^{i-k}, \quad
B_j=\sum_{l=0}^{j}\frac{d_{l}}{(j-l)!}(q\partial_{q})^{j-l},\\
\xi=&pz_1+p^2z_2+p^3z_3+\xi_0, \quad \eta=qz_1-q^2z_2+q^3z_3+\eta_0.
\end{aligned}
\end{equation}
For simplicity, we can rewrite the functions $M^{(n)}_{i,j}$ as
\begin{equation}
\begin{aligned}
M^{(n)}_{i,j}&=\sum_{k=0}^{i}\frac{c_{k}}{(i-k)!}(p\partial_{p}+\xi^{'}
+n)^{i-k}\\
\times & \sum_{l=0}^{j}\frac{d_{l}}{(j-l)!}(q
\partial_{q}+\eta^{'}-n)^{j-l}\frac{1}{p+q},
\end{aligned}
\end{equation}
where
\begin{equation}
\begin{aligned}
\xi^{'}=&pz_1+2p^2z_2+3p^3z_3, \quad \eta^{'}=qz_1-2q^2z_2+3q^3z_3,
\end{aligned}
\end{equation}
and $p$, $q$, $c_k$, and $d_l$ are arbitrary complex constants. Further,
taking the parameter constraints
$$p=q=1,\qquad c_k=d^*_k,$$
setting $\tau^{'}_{0}=f$, $z_{1}=k_1x_1+k_2x_2$, $z_{2}=\sqrt{\frac{{k_1k_2}
(k^2_2-k^2_1)}{2k_3k_4}}i(k_3y_1+k_4y_2)$, and
 $z_{3}=-2k_2(k^2_2-k^2_1)t$, the rational solutions of the 4D Fokas equation can be
generated from equation \eqref{kphierarchy}.
Based on the above results, the rational solutions of the 4D Fokas equation are
presented in the following Theorem.

\textbf{Theorem 2.}   The $(4 + 1)$-dimensional Fokas equation \eqref{N1} has
 rational solutions
\begin{equation}\label{rational}
\begin{aligned}
u=(k^2_2-k^2_1)(\ln f)_{xx},\quad x=k_1x_1+k_2x_2,\\
\end{aligned}
\end{equation}
where
\begin{equation}\label{r1}
\begin{aligned}
f= \det\limits_{1\leq i,j\leq N}(M^{(n)}_{2i-1,2j-1})\mid_{n=0}.\\
\end{aligned}
\end{equation}
The matrix elements in $f$ are defined by
\begin{equation}\label{r2}
\begin{aligned}
M_{i,j}^{(n)}&=\sum_{k=0}^{i}\frac{c_{k}}{(i-k)!}(p\partial_{p}+\xi^{'}+n)^{i-k}\\
&\times \sum_{l=0}^{j}\frac{c_{l}^{*}}{(j-l)!}(p^{*}
\partial_{p^{*}}+\xi^{'*}-n)^{j-l}\frac{1}{p+p^{*}}\mid_{p=1},\\
\end{aligned}
\end{equation}
\begin{equation*}
\begin{aligned}
\xi^{'}&=k_1x_1+k_2x_2+\sqrt{\frac{{2k_1k_2k_3}
(k^2_2-k^2_1)}{-k_4}}y_1\\
&+\sqrt{\frac{{2k_1k_2k_4}
(k^2_2-k^2_1)}{-k_3}}y_2-3k_2(k^2_2-k^2_1)t,
\end{aligned}
\end{equation*}
the asterisk denotes the complex conjugation,
$i, j, k$, and $l$ are arbitrary positive integers, and  $c_{k}$
and $c_{l}$  are arbitrary complex constants. We take
$k_1=1, k_2=\frac{6}{5}, k_3=1$, and $k_4=1$ in this section.

\subsection{Fundamental rational solution}

\begin{figure}[!htb]
\centering
\subfigure[t=0]{\includegraphics[height=3cm,width=5cm]{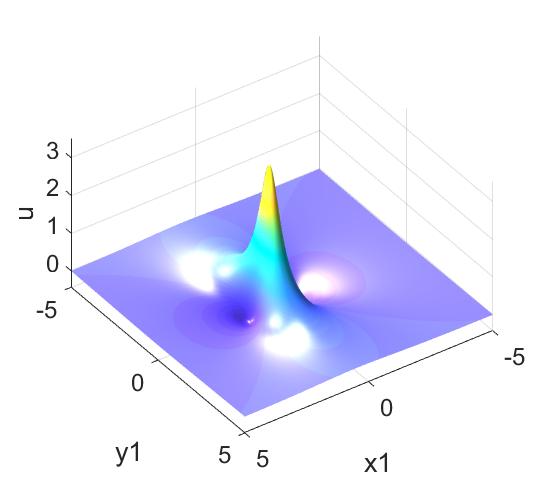}}\quad
\subfigure[t=0]{\includegraphics[height=3cm,width=5cm]{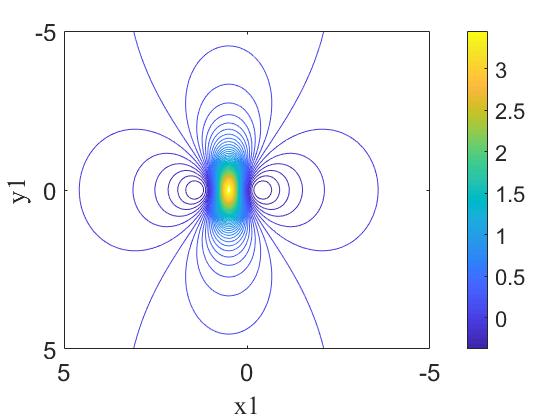}}
\caption{ First-order rational solution for the 4D Fokas equation in the $(x_1,y_1)$-plane
with parameters $N=1, c_0=1, c_1=0, k_1=1, k_2=\frac{6}{5}, k_3=1, k_4=1, x_2=0, y_2=0$,
and $t=0$ in equation \eqref{rational}.
Panel (b) is the contour plot of panel (a).}
\label{rational-1}
\end{figure}

According to Theorem 2, taking the parameters $N=1$, $c_0=1$,
and $c_1=0$ in equation \eqref{rational}, we first derive
the fundamental rational solution of the 4D Fokas equation:
\begin{equation}
\begin{aligned}
u=2(k^2_1-k^2_2)k_3k_4\frac{2k_1k_2(k^2_1-k^2_2)[k_3y_1+k_4y_2]^2
+k_3k_4l_{lump}^2-\frac{1}{4}k_3k_4}
{\left (18k_1k_2(k^2_2-k^2_1)[k_3y_1+k_4y_2]^2
+9k_3k_4l_{lump}^2+\frac{9}{4}k_3k_4\right)^2},
\end{aligned}
\end{equation}
where
\begin{equation}
\begin{aligned}
l_{lump}=k_1x_1+k_2x_2+3(k^2_1k_2-k^3_2)t
+\frac{1}{2}.
\end{aligned}
\end{equation}
As can be seen from the above expressions, in order to
ensure that the fundamental rational solution
is non-singular, $k_1k_2k_3k_4(k_2^2-k_1^2)>0$
must be held. The fundamental rational solution
is a lump and has the following extreme points
in the $(x_1,y_1)$-plane, see Fig. \ref{rational-1}:
\begin{equation*}
\begin{aligned}
\Lambda_1&=(x_{11},y_{11})=\left(-\frac{k_2}{k_1}x_2
+3(\frac{k^3_2}{k_1}-k_1k_2)t+\frac{1}{2k_1},y1=-\frac{k_4}{k_3}y_2\right),\\
\Lambda_2&=(x_{12},y_{12})=\left(-\frac{k_2}{k_1}x_2
+3(\frac{k^3_2}{k_1}-k_1k_2)t+\frac{1+\sqrt{3}}{2k_1},y1=-\frac{k_4}{k_3}y_2\right),\\
\Lambda_3&=(x_{13},y_{13})=\left(-\frac{k_2}{k_1}x_2
+3(\frac{k^3_2}{k_1}-k_1k_2)t+\frac{1-\sqrt{3}}{2k_1},y1=-\frac{k_4}{k_3}y_2\right)\textcolor{red}{.}
\end{aligned}
\end{equation*}
After simple calculations, we get a maximum value $H_{Max}=H(x_{11},y_{11})=8(k^2_2-k^2_1)$
and two minimum values $H_{Min}=H(x_{12},y_{12})=H(x_{13},y_{13})=k^2_1-k^2_2$ of the lump solution.
 The lump trajectory  is $k_1x_1+k_2x_2+3(k^2_1k_2-k^3_2)t
+\frac{1}{2}=0$.

\subsection{High-order rational solutions}

\begin{figure}[!htb]
\centering
\subfigure[t=0]{\includegraphics[height=3cm,width=5cm]{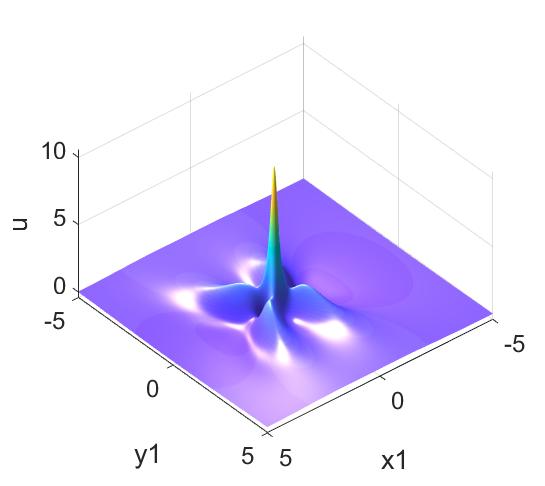}}\quad
\subfigure[t=0]{\includegraphics[height=3cm,width=5cm]{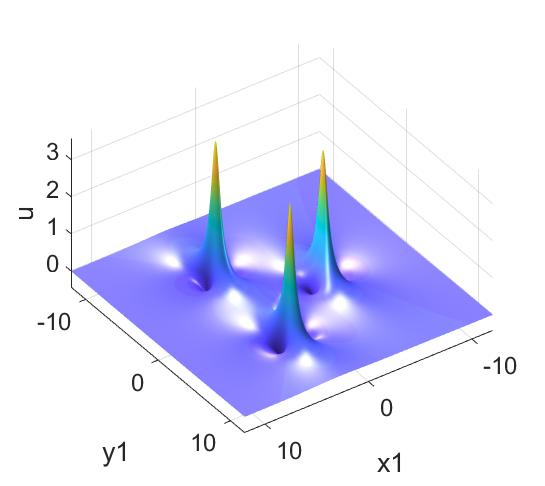}}\\
\subfigure[t=0]{\includegraphics[height=3cm,width=5cm]{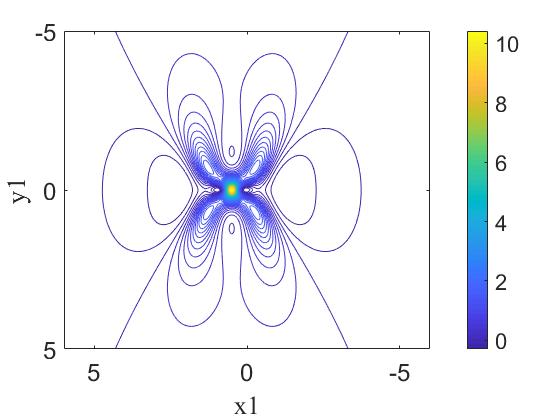}}\quad
\subfigure[t=0]{\includegraphics[height=3cm,width=5cm]{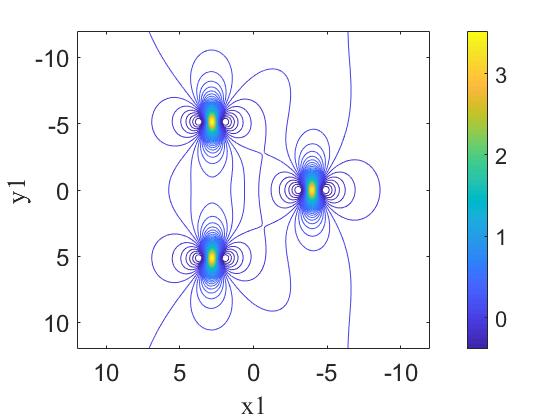}}
\caption{Second-order rational solutions of the 4D Fokas equation
with parameters $N=2,  c_1=0, c_2=0, k_1=1, k_2=\frac{6}{5}
,k_3=1, k_4=1, x_2=0, y_2=0$, and $t=0$ in equation \eqref{rational}.
(a): a fundamental pattern with $c_0=1$ and $c_3=-\frac{1}{12}$; (b):
a triangular pattern with $c_0=1$ and $c_3=-30$. Panels (c)
and (d) are the contour plots of panels (a) and (b), respectively.}
\label{rational-2}
\end{figure}

\begin{figure}[!htb]
\centering
\subfigure[t=0]{\includegraphics[height=3cm,width=3.8cm]{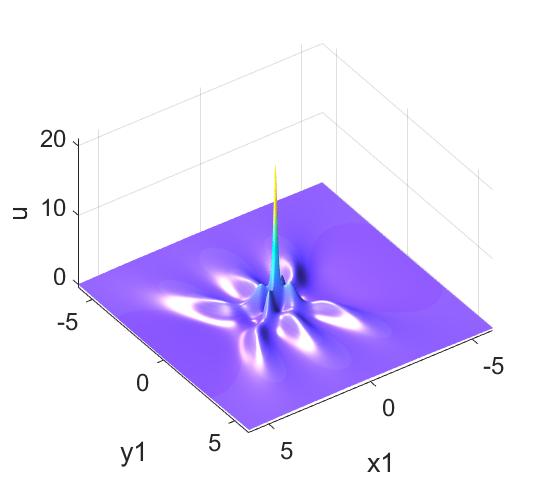}}
\subfigure[t=0]{\includegraphics[height=3cm,width=3.8cm]{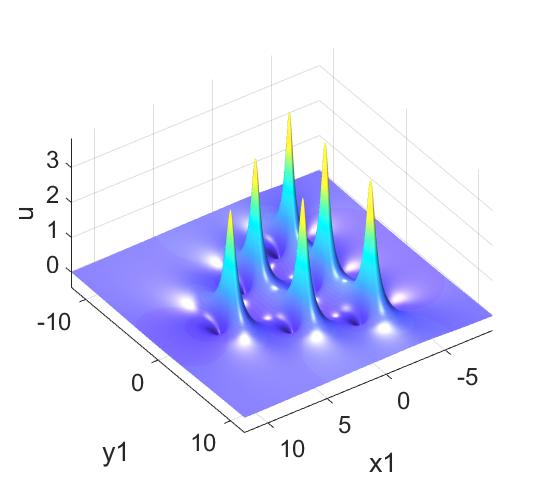}}
\subfigure[t=0]{\includegraphics[height=3cm,width=3.8cm]{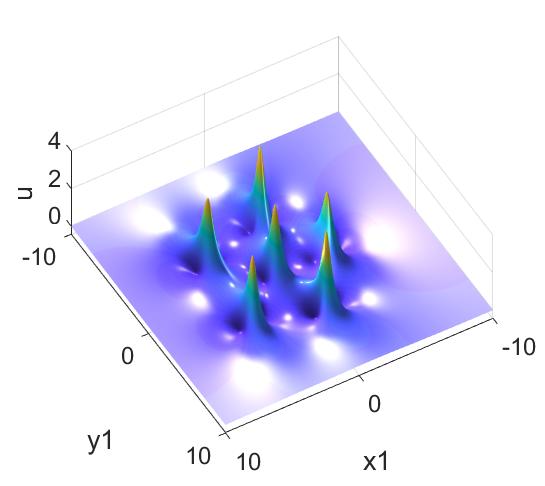}}
\subfigure[t=0]{\includegraphics[height=3cm,width=3.8cm]{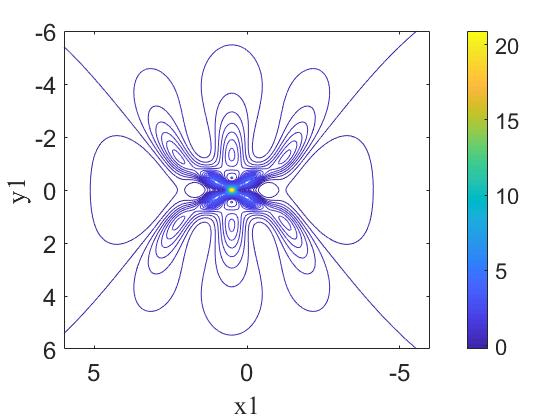}}
\subfigure[t=0]{\includegraphics[height=3cm,width=3.8cm]{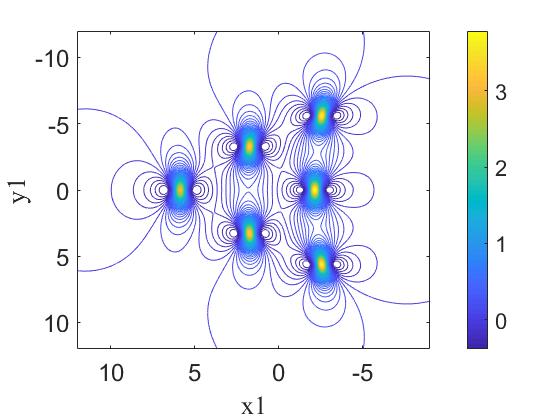}}
\subfigure[t=0]{\includegraphics[height=3cm,width=3.8cm]{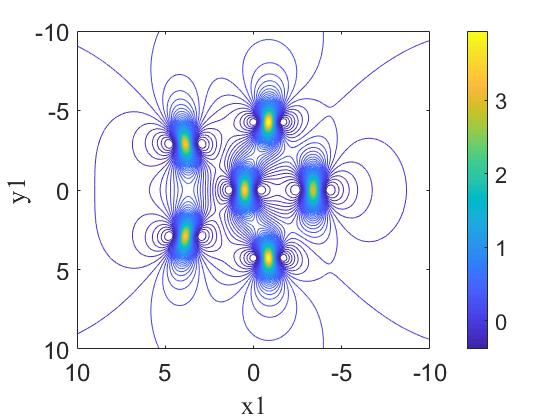}}
\caption{ Third-order rational solutions of the 4D Fokas equation
with parameters $N=3,  c_1=0, c_2=0, c_4=0, k_1=1, k_2=\frac{6}{5}
,k_3=1, k_4=1, x_2=0, y_2=0$, and $t=0$ in equation \eqref{rational}.
(a): a fundamental pattern with $c_0=1, c_3=-\frac{1}{12}$ and $c_3=-\frac{1}{240}$; (b):
a triangular pattern with $c_0=1, c_3=-\frac{25}{3}$ and $c_5=0$;
(c): a ring pattern with $c_0=1, c_3=0$ and $c_5=20$.
Panels (d), (e), and (f) are the contour plots of panels (a), (b), and (c), respectively.}
\label{rational-3}
\end{figure}

In this section, we consider the high-order rational
solutions of the 4D Fokas equation. $N$-order lump solutions
are derived in the $(x_1,y_1)$-plane from Theorem 2 for any given $N$.
For example, taking $N=2, k_1=1, k_2=\frac{6}{5}
,k_3=1$, and $k_4=1$, the second-order lump solutions $u_{\rm 2ord}$
are obtained
\begin{equation}\label{2-lump}
u_{\rm 2ord}=(k_2^2-k^2_1) \left (\ln\begin{vmatrix}
M^{(0)}_{11} & M^{(0)}_{13} \\ M^{(0)}_{31} & M^{(0)}_{33} \end{vmatrix}\right )_{xx} ,
\end{equation}
where $M_{i,j}^{(n)}$ are defined in Theorem 2. As shown in Fig. \ref{rational-2},
the second-order lumps have two types of patterns, which are controlled by four free parameters.
Similarly, the third-order lump solutions
are derived by taking $N=3, k_1=1, k_2=\frac{6}{5}
,k_3=1$, and $k_4=1$.  The third-order lumps have three types of patterns, which are
controlled by six parameters, see Fig. \ref{rational-3}.
For larger values of $N$, as more free parameters will be generated,
the patterns of the lumps will be more abundant and their dynamic
behavior will be more complicated. We note that the pattern dynamics
of high-order lumps is similar to rogue waves dynamics in
 $(1+1)$-dimensional systems.

\section{Semi-rational solutions in the determinant form} \label{4}

In this section, we present a Theorem for constructing the
semi-rational solutions of the 4D Fokas equation.
In order to obtain the semi-rational solutions of the 4D Fokas equation,
we will first introduce the following differential operators
\begin{equation}
\begin{aligned}
\Xi_i=\sum_{k=0}^{n_i}a_{ik}(p_i\partial_{p_i})^{n_i-k}, \quad
\mho_j=\sum_{l=0}^{n_j}a^*_{jl}(p^*_j\partial_{p^*_j})^{n_j-l}.\\
\end{aligned}
\end{equation}
We choose the following functions
\begin{equation}
\begin{aligned}
&\varphi^{(n)}_{i}=\Xi_ip^{n}_{i}e^{\xi^{i}},\\
&\psi^{(n)}_{j}=\mho_j(-q_{j})^{-n}e^{\eta_{j}},\\
&K^{(n)}_{i,j}=\Xi_i\mho_j\frac{1}{p_{i}+p^*_{j}}[\delta_{ij}
+(-\frac{p_i}{p^*_j})^ne^{\xi_i+\xi^*_j}].\\
\end{aligned}
\end{equation}
The functions $\varphi^{(n)}_{i}$ and $\psi^{(n)}_{j}$
also satisfy the equation \eqref{lemma1}. For simplicity, we rewrite the matrix element $K^{(n)}_{i,j}$ as
\begin{equation}
\begin{aligned}
K_{i,j}^{(n)}&=(-\frac{p_i}{p^*_j})e^{\xi_i+\xi^*_j}\sum_{k=0}^{n_i}a_{ik}
(p_i\partial_{p_i}+\xi^{'}_i+n)^{n_i-k}\\
& \times \sum_{l=0}^{n_j}a_{jl}^{*}(p^{*}_j
\partial_{p^{*}_j}+\xi^{'*}_j-n)^{n_j-l}\frac{1}{p_i+p^{*}_j}+\delta_{ij}a_{in_i}a_{jn_j}^{*},
\end{aligned}
\end{equation}
where
\begin{equation}
\begin{aligned}
\xi_i&=p_iz_1+p_i^2z_2+p_i^3z_3+\xi_{i0},\\
\xi^{'}_i&=p_iz_1+2p_i^2z_2+3p_i^3z_3.
\end{aligned}
\end{equation}
Here $p_i$ and $a_{ik}$ are arbitrary complex
constants,  $\delta_{ij}=0,1$, and $n_i$ are arbitrary
positive integers. Furthermore, taking $\tau_0=f$,
$z_{1}=k_1x_1+k_2x_2$, $z_{2}=\sqrt{\frac{{k_1k_2}
(k^2_2-k^2_1)}{2k_3k_4}}i(k_3y_1+k_4y_2)$, and
$z_{3}=-2k_2(k^2_2-k^2_1)t$, then, the semi-rational solutions
of the 4D Fokas equation would be derived.
Thus, semi-rational solutions
of 4D Fokas equation can be determined by the
following Theorem.

\textbf{Theorem 3.}   The $(4 + 1)$-dimensional Fokas equation \eqref{N1} has
 semi-rational solutions
\begin{equation}\label{BBT1}
\begin{aligned}
u=(k^2_2-k^2_1)(\ln f)_{xx},\quad x=k_1x_1+k_2x_2,\\
\end{aligned}
\end{equation}
where
\begin{equation}\label{Bt1}
\begin{aligned}
f= \det\limits_{1\leq i,j\leq N}(K^{(n)}_{i,j})\mid_{n=0},\\
\end{aligned}
\end{equation}
and the matrix elements in $f$ are defined by
\begin{equation}\label{mm1}
\begin{aligned}
K_{i,j}^{(n)}&=(-\frac{p_i}{p^*_j})e^{\xi_i+\xi^*_j}
\sum_{k=0}^{n_i}c_{ik}(p_i\partial_{p_i}+\xi^{'}_i+n)^{n_i-k}\\
 \times &\sum_{l=0}^{n_j}c_{jl}^{*}(p^{*}_j
\partial_{p^{*}_j}+\xi^{'*}_j-n)^{n_j-l}\frac{1}{p_i+p^{*}_j}+\delta_{ij}a_{in_i}a_{jn_j}^{*},
\end{aligned}
\end{equation}
\begin{equation*}
\begin{aligned}
\xi_i&=k_1p_ix_1+k_2p_ix_2+\sqrt{\frac{{k_1k_2k_3}
(k^2_2-k^2_1)}{-2k_4}}p_i^2y_1\\
&+\sqrt{\frac{{k_1k_2k_4}(k^2_2-k^2_1)}{-2k_3}}p_i^2y_2-k_2(k^2_2-k^2_1)p_i^3t+\xi_{i0},\\
\xi^{'}_i&=k_1p_ix_1+k_2p_ix_2+\sqrt{\frac{{2k_1k_2k_3}
(k^2_2-k^2_1)}{-k_4}}p^2_iy_1\\
&+\sqrt{\frac{{2k_1k_2k_4}(k^2_2-k^2_1)}{-k_3}}p^2_iy_2-3k_2(k^2_2-k^2_1)p^3_it.
\end{aligned}
\end{equation*}
The asterisk denotes the complex conjugation, $i,j,k$, and $l$
are arbitrary positive integers,  and $k_1$, $k_2$, $k_3$, and
$k_4$ are arbitrary real constants. It is not difficult to find
that the semi-rational solutions will become rational
solutions when $\delta_{ij}=0$. We note that the patterns of rational solutions
are similar to those corresponding to rational solutions of the Davey-Stewartson equation, reported in
Refs. \cite{rao1,rao2,qc}.

\subsection{Lumps on one-soliton background}

\begin{figure}[!htb]
\centering
\subfigure[t=-5]{\includegraphics[height=2cm,width=3.8cm]{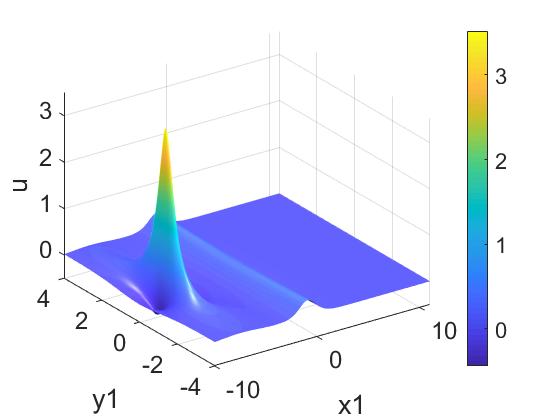}}
\subfigure[t=0]{\includegraphics[height=2cm,width=3.8cm]{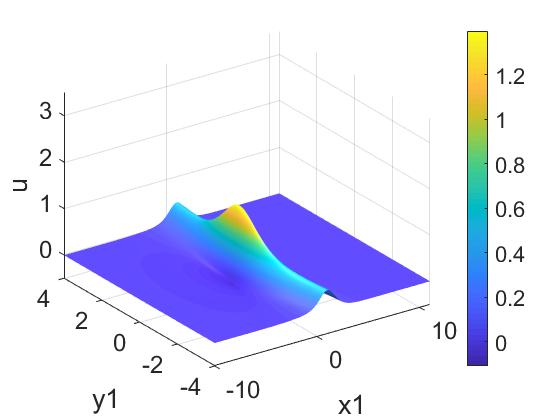}}
\subfigure[t=5]{\includegraphics[height=2cm,width=3.8cm]{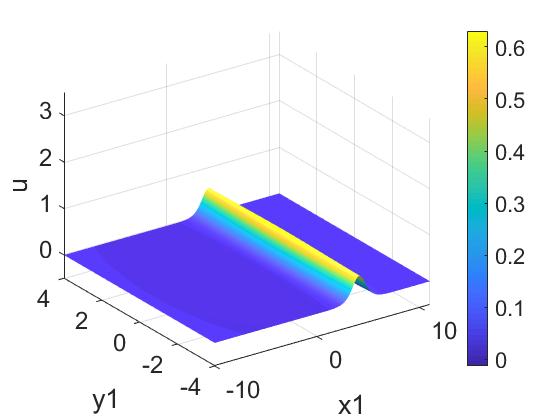}}\\
\subfigure[t=-5]{\includegraphics[height=2cm,width=3.8cm]{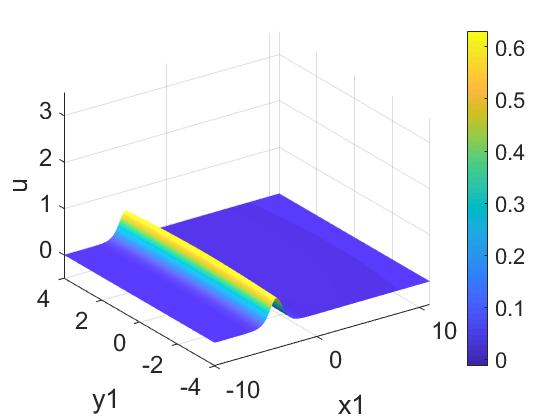}}
\subfigure[t=0]{\includegraphics[height=2cm,width=3.8cm]{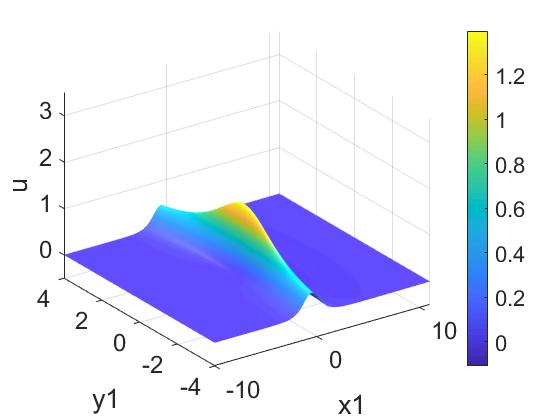}}
\subfigure[t=5]{\includegraphics[height=2cm,width=3.8cm]{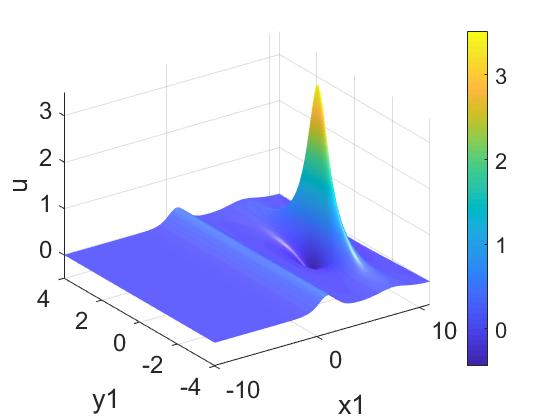}}\\
\subfigure[t=-5]{\includegraphics[height=2cm,width=3.8cm]{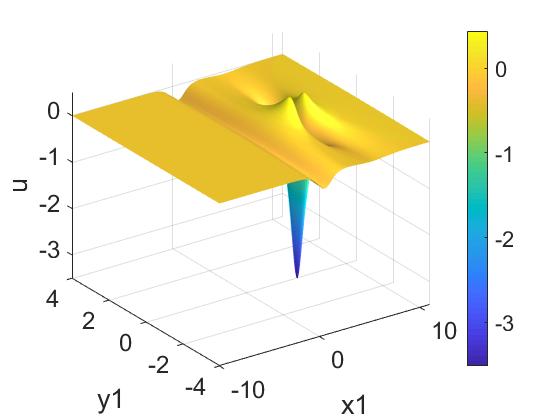}}
\subfigure[t=0]{\includegraphics[height=2cm,width=3.8cm]{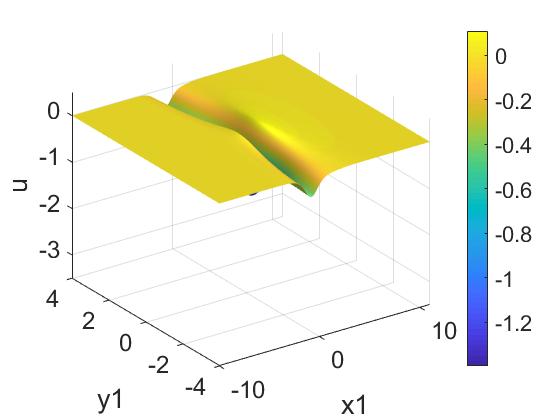}}
\subfigure[t=5]{\includegraphics[height=2cm,width=3.8cm]{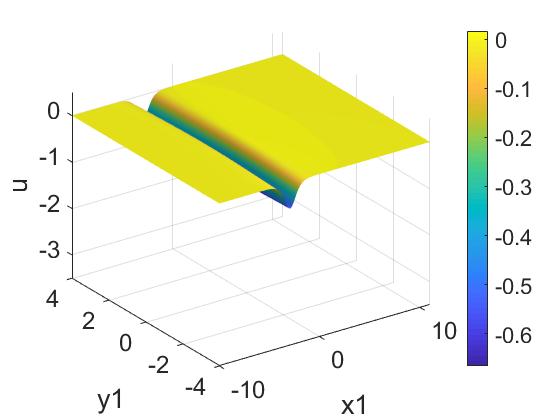}}\\
\subfigure[t=-5]{\includegraphics[height=2cm,width=3.8cm]{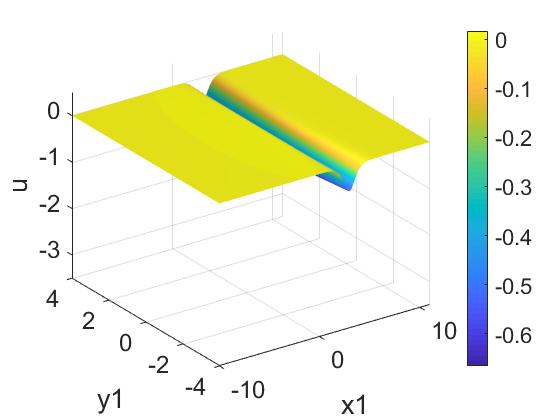}}
\subfigure[t=0]{\includegraphics[height=2cm,width=3.8cm]{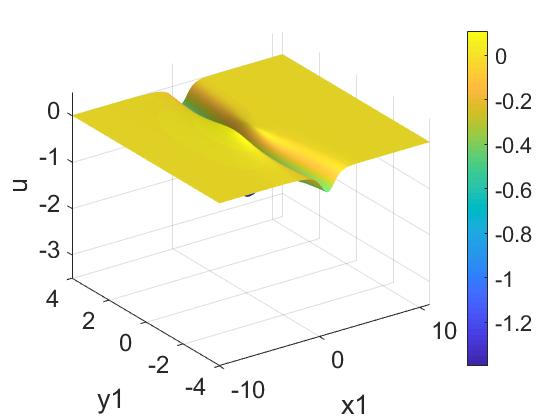}}
\subfigure[t=5]{\includegraphics[height=2cm,width=3.8cm]{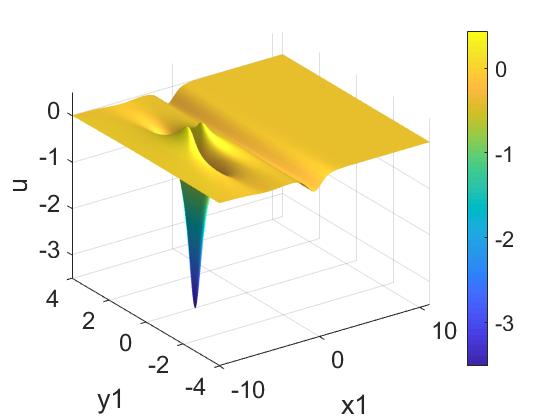}}
\caption{  Time evolution of semi-rational solution $u_{\rm ls}$ of the 4D Fokas equation,
(a,b,c): A bright lump is annihilated by a bright soliton with parameters $k_1=-1,k_2=-1.2,k_3=1,k_4=1, x_2=0$, and $y_2=0$;
(d,e,f): A bright lump is created from a bright soliton with parameters
$k_1=1,k_2=-1.2,k_3=1,k_4=1, x_2=0$, and $y_2=0$;
(g,h,i): A dark lump is annihilated by a dark soliton with parameters
$k_1=-1.2,k_2=1,k_3=1,k_4=-1, x_2=0$, and $y_2=0$;
(j,k,l): A dark lump is created from a dark soliton with parameters
$k_1=-1.2,k_2=-1,k_3=1,k_4=-1, x_2=0$, and $y_2=0$.}
\label{1lump-1soliton}
\end{figure}

The semi-rational solution $u_{\rm ls}$ consisting of a lump
and a soliton is derived by taking
$N=1,n_i=1, a_{01}=0, a_{11}=1, p_1=1,
\delta_{ii}=1$, and $\delta_{ij}=0 \,(i\neq j)$
in equation \eqref{BBT1}.
The expression of $u_{\rm ls}$ is as follows
\begin{equation}
\begin{aligned}
u_{\rm ls}=2(k^2_1-k^2_2)k_3k_4\frac{\begin{Bmatrix}& [2k_1k_2(k^2_1-k^2_2)l_1^2
+k_3k_4l_{lump}^2-\frac{1}{4}k_3k_4]e^{\xi}\\
&-4k_3k_4(l_{lump}+1)^2
+8k_1k_2(k^2_1-k^2_2)l_1^2+k_3k_4\end{Bmatrix}e^{\xi}}
{\left ([18k_1k_2(k^2_2-k^2_1)l_1^2+9k_3k_4l_{lump}^2+\frac{9}{4}k_3k_4]e^{\xi}+18k_3k_4\right)^2},
\end{aligned}
\end{equation}
where
$$l_1=k_3y_1+k_4y_2,\quad \xi=2k_1x_1+2k_2x_2+2(k_1^2k_2-k_2^3)t,$$
$$l_{lump}=k_1x_1+k_2x_2+3(k^2_1k_2-k^3_2)t.$$
By choosing different parameters $k_1, k_2, k_3$. and $k_4$, we derive
a lump fusing into or fissioning from
a dark soliton or from a bright soliton, see Fig. \ref{1lump-1soliton}.
The classification of four different types of interaction between lumps and one-soliton solutions
is given in Table 1. From the above results,
we can easily calculate that the velocities of lump and soliton
are $V_{\rm lump}=3\frac{k^2_1k_2-k^3_2}{k_1}$ and
$V_{\rm soliton}=\frac{k^2_1k_2-k^3_2}{k_1}$, respectively.
The velocity of lump is always greater than that of soliton,
see Fig. \ref{1lump-1soliton}.

\begin{table}
\centering
\begin{tabular}{ccc}
\hline
Parameter condition (i)&\quad \quad \quad Parameter condition (ii)&\quad \quad \quad \quad Results\\
\hline
$k_3k_4>0$&\quad \quad \quad $k_2<k_1<0$&\quad \quad \quad \quad a lump is annihilated\\
$k_3k_4<0$&\quad \quad \quad $|k_2|>k_1>0>k_2$&\quad \quad \quad \quad  by a bright soliton\\
\cline{1-3}
$k_3k_4>0$&\quad \quad \quad $k_2>k_1>0$&\quad \quad \quad \quad a lump is created\\
$k_3k_4<0$&\quad \quad \quad $k_2>|k_1|>0>k_1$&\quad \quad \quad \quad from a bright soliton\\
\cline{1-3}
$k_3k_4>0$&\quad \quad \quad $|k_1|>k_2>0>k_1$&\quad \quad \quad \quad a lump is annihilated\\
$k_3k_4<0$&\quad \quad \quad $k_1>k_2>0$&\quad \quad \quad \quad by a dark soliton\\
\cline{1-3}
$k_3k_4>0$&\quad \quad \quad $k_1>|k_2|>0>k_2$&\quad \quad \quad \quad a lump is created\\
$k_3k_4<0$&\quad \quad \quad $k_1<k_2<0$&\quad \quad \quad \quad from a dark soliton\\
\cline{1-3}
\end{tabular}
\caption{Classification of four different types of interactions between lumps  and one-soliton solution.}
\end{table}

\begin{figure}[!htb]
\centering
\subfigure[t=-7]{\includegraphics[height=3cm,width=4cm]{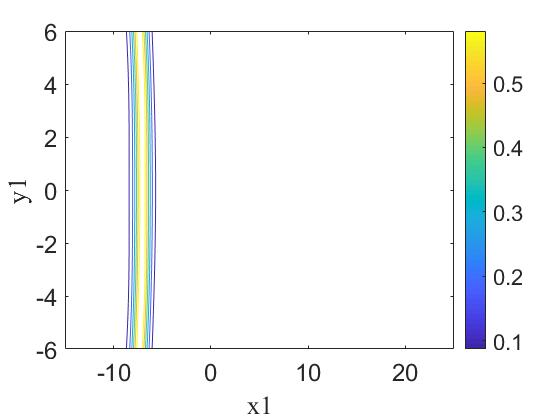}}
\subfigure[t=-1]{\includegraphics[height=3cm,width=4cm]{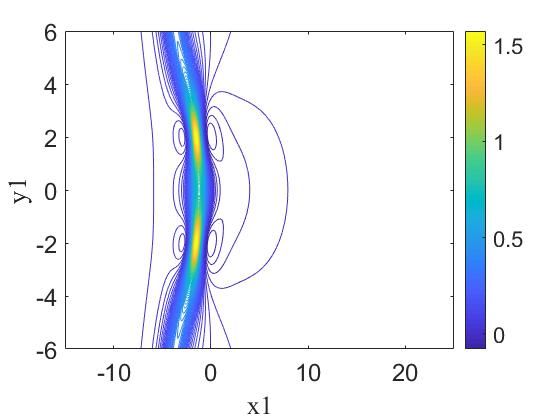}}
\subfigure[t=1]{\includegraphics[height=3cm,width=4cm]{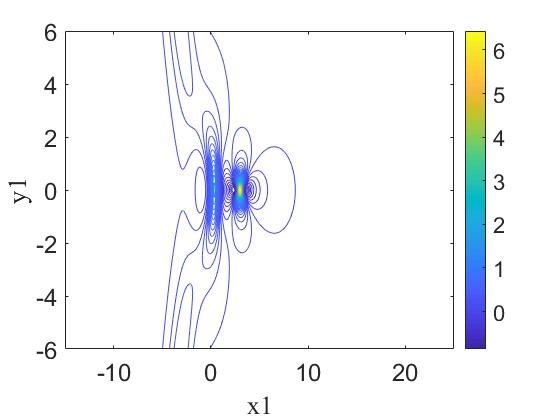}}
\subfigure[t=7]{\includegraphics[height=3cm,width=4cm]{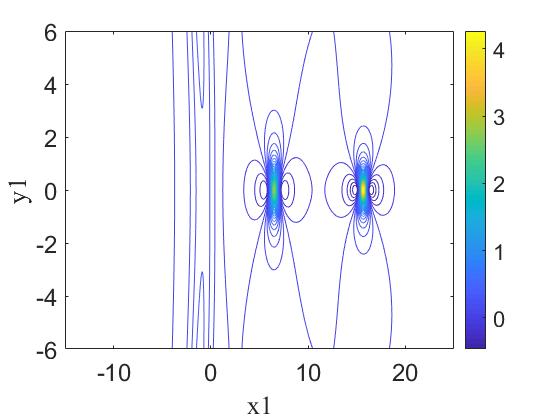}}
\caption{ Time evolution of the process of fission of two lumps
from a bright soliton of the 4D Fokas equation with parameters $N=1,n_i=2, a_{01}=0,
a_{11}=4, a_{12}=1, p_1=1, \delta_{ii}=1, \delta_{12}=0,
k_1=1, k_2=1.2, k_3=1, k_4=1, x_2=0$, and $y_2=0$ in equation \eqref{BBT1}.}
\label{2lump-1soliton}
\end{figure}

The semi-rational solutions consisting of more lumps and
a soliton are generated for $N=1$ and $n_i\geq2$ in equation \eqref{BBT1}.
For example, taking
$N=1,n_i=2, a_{01}=0, a_{11}=4, a_{12}=1, p_1=1,
\delta_{11}=1$,  and $\delta_{12}=0$
in equation \eqref{BBT1}, we can derive the semi-rational
solutions composed of two lumps and a soliton.
There are also four distinct types of such semi-rational solutions.
To have an idea of the dynamics of such semi-rational solutions, we show here only the process
of fission of two lumps from a bright soliton, see Fig. \ref{2lump-1soliton}.

\subsection{Lumps on multi-solitons background}

\begin{figure}[!htb]
\centering
\subfigure[t=-10]{\includegraphics[height=3cm,width=3.8cm]{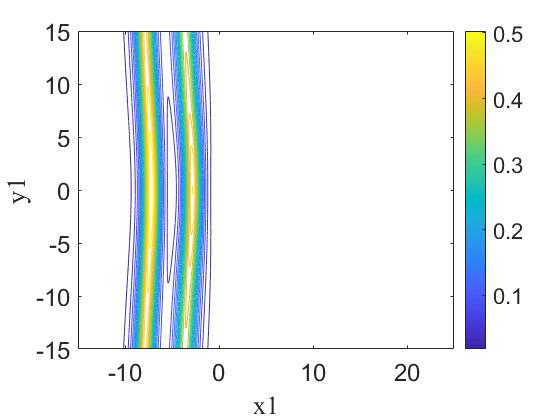}}
\subfigure[t=-2]{\includegraphics[height=3cm,width=3.8cm]{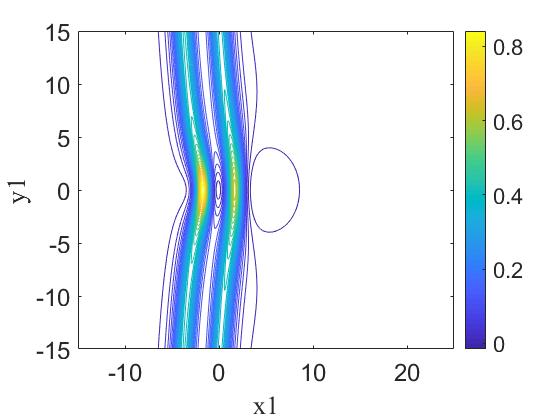}}
\subfigure[t=0]{\includegraphics[height=3cm,width=3.8cm]{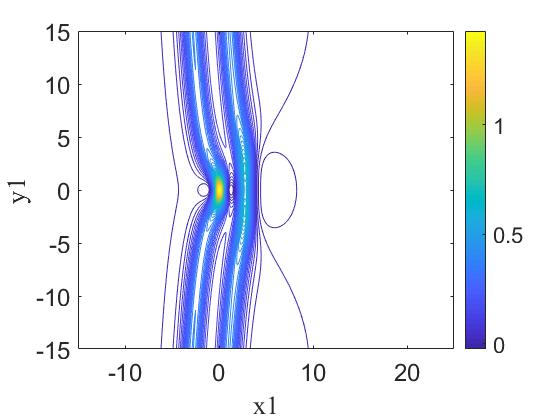}}
\subfigure[t=2]{\includegraphics[height=3cm,width=3.8cm]{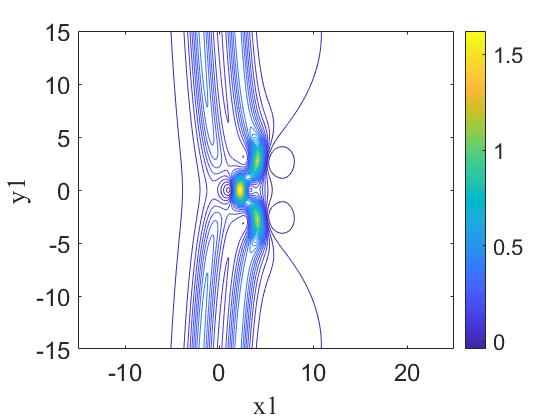}}\quad
\subfigure[t=10]{\includegraphics[height=3cm,width=3.8cm]{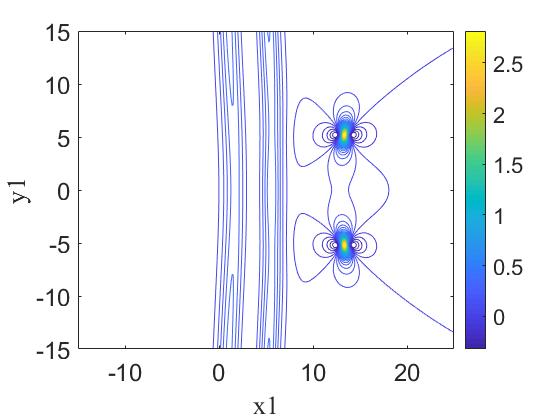}}
\caption{The time evolution of fission of two lumps
from two solitons for the 4D Fokas equation with parameters $N=1, n_i=2, a_{10}=1, a_{20}=1, a_{11}=1, a_{21}=1,
p_1=\frac{1}{2}, p_2=\frac{1}{3}, \delta_{11}=1, \delta_{12}=0, \delta_{21}=0, \delta_{22}=1,
k_1=1, k_2=2, k_3=1, k_4=1, x_2=0$, and $y_2=0$ in equation \eqref{BBT1}.}
\label{2lump-2soliton}
\end{figure}

For $N\geq2, n_i=1, \delta_{11}=1, \delta_{12}=0, \delta_{21}=0$, and $\delta_{22}=1$  in equation \eqref{BBT1}, the semi-rational
solutions consisting of more lumps and more solitons are derived.
For example, taking $N=1, n_i=2, k_1=1, k_2=2, k_3=1, k_4=1, a_{10}=1,
a_{20}=1, a_{11}=1, a_{21}=1, p_1=\frac{1}{2}, p_2=\frac{1}{3},
\delta_{11}=1, \delta_{12}=0, \delta_{21}=0$, and $\delta_{22}=1$
in equation \eqref{BBT1}, we obtain the interaction of local wave structures
described by two lumps and two solitons. The exact
expression of the corresponding solution $u_{2ls}$ is as follows
\begin{equation}
\begin{aligned}
u_{\rm 2ls}=(k_2^2-k^2_1) \left (\ln\begin{vmatrix}
K^{(0)}_{11} & K^{(0)}_{12} \\ K^{(0)}_{21} & K^{(0)}_{22} \end{vmatrix}\right )_{xx} ,
\end{aligned}
\end{equation}
where
$$K^{(0)}_{11}=1+\left(\frac{3}{4}(y_1+y_2)^2+\frac{1}{4}(x_1+2x_2-\frac{9}{2}t+1)^2
+\frac{1}{4}\right)e^{x_1+2x_2-\frac{3}{2}t},$$
$$K^{(0)}_{12}=\left(\frac{2(y_1+y_2)^2}{5}+\frac{1}{5}(x_1+2x_2-\frac{13}{4}t+\frac{13}{10})^2
-\frac{5}{16}(t+\frac{2}{5})^2+\frac{36}{125}+A\right)e^{\varrho_1},$$
$$K^{(0)}_{21}=\left(\frac{2(y_1+y_2)^2}{5}+\frac{1}{5}(x_1+2x_2-\frac{13}{4}t+\frac{13}{10})^2
-\frac{5}{16}(t+\frac{2}{5})^2+\frac{36}{125}-A\right)e^{\varrho_2},$$
$$K^{(0)}_{22}=1+\left(\frac{2}{9}(y_1+y_2)^2+6(x_1+2x_2-2t+\frac{3}{2})^2
+\frac{3}{8}\right)e^{\frac{2}{3}x_1+\frac{4}{3}x_2-\frac{4}{9}t},$$
$$A=\sqrt{-3}(y_1+y_2)(\frac{1}{3}x_1+\frac{2}{3}x_2+t+\frac{9}{15}),$$
$$\varrho_1=\frac{5}{6}x_1+\frac{5}{3}x_2-\frac{35}{36}t+\frac{5\sqrt{-3}}{36}(y_1+y_2),$$
$$\varrho_2=\frac{5}{6}x_1+\frac{5}{3}x_2-\frac{35}{36}t-\frac{5\sqrt{-3}}{36}(y_1+y_2).$$
The five panels in Fig. \ref{2lump-2soliton} describe the process of creation of two lumps from
the background of two solitons. As shown in Fig. \ref{2lump-2soliton}, with time evolution
more peaks are created during the interaction between lumps and solitons around $t=2.5$,
then two lumps and two solitons are completely separated around $t=10$.

\subsection{A hybrid of two lumps, a breather, and a soliton}

\begin{figure}[!htb]
\centering
\subfigure[t=-4]{\includegraphics[height=3.5cm,width=5.5cm]{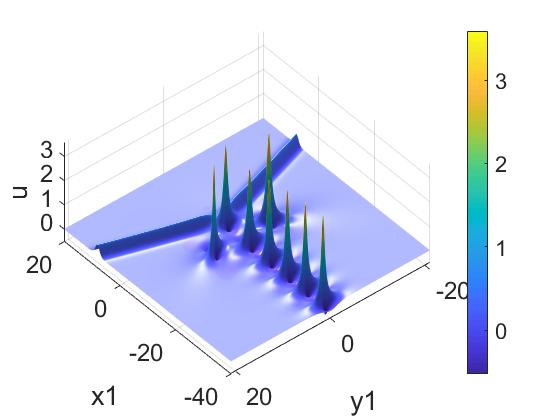}}\quad
\subfigure[t=4]{\includegraphics[height=3.5cm,width=5.5cm]{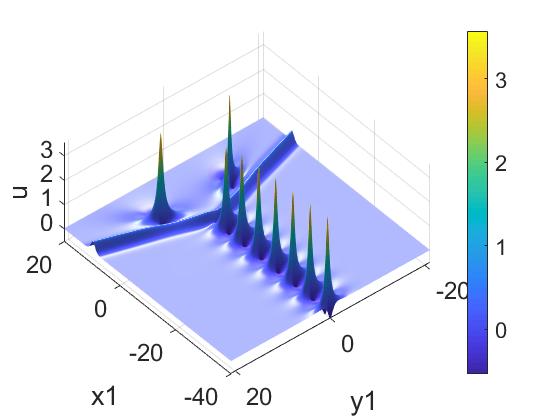}}
\caption{ Time evolution of semi-rational solution consisting of two lumps, a breather, and a V-type soliton of
the 4D Fokas equation with parameters $N=2, n_i=2, k_1=1, k_2=2, k_3=1, k_4=1,
a_{10}=1, a_{20}=1, a_{11}=1, a_{21}=1, p_1=1+\frac{i}{2},
p_2=1-\frac{i}{2}, \delta_{11}=1, \delta_{12}=1, \delta_{21}=1,
\delta_{22}=1, x_2=0$, and $y_2=0$ in equation \eqref{BBT1}.}
\label{2lump-1soliton-bre}
\end{figure}

The third type of semi-rational solution consisting of two lumps,
a breather, and a V-type soliton is derived for $N=2, n_i=2$ and
$\delta_{ij}=1$ in equation \eqref{BBT1}. The corresponding semi-rational
solution is shown in Fig. \ref{2lump-1soliton-bre}, Obviously, the
process of their interaction is elastic,
the amplitudes and shapes of soliton, breather, and lumps did not change after the interaction.
This type of semi-rational solution has
never been reported elsewhere, to the best of our knowledge.

\section{Rational and semi-rational solutions related to the 3D KP equation}\label{5}

The 3D KP equation can be read as follows \cite{PRL}
\begin{equation}\label{3dkp1}
\begin{aligned}
\left(W_\varsigma+6WW_\upsilon+W_{\upsilon\upsilon\upsilon}\right)_\upsilon
-\alpha \left(W_{\varphi\varphi}+W_{ZZ}\right)=0.
\end{aligned}
\end{equation}
It describes the dynamic behavior of nonlinear waves
and solitons in plasma and fluids \cite{3d-1,3d-2}. The rational
and semi-rational solutions of the 3D KP equation can be expressed
in Theorems 4 and 5 as follows.

\textbf{Theorem 4.}   The 3D KP equation \eqref{3dkp1} has
 rational solutions
\begin{equation}\label{3dkp-rational}
\begin{aligned}
W=2(\ln f)_{\upsilon\upsilon},\\
\end{aligned}
\end{equation}
where
\begin{equation}\label{3dr1}
\begin{aligned}
f= \det\limits_{1\leq i,j\leq N}(H  ^{(n)}_{2i-1,2j-1})\mid_{n=0},\\
\end{aligned}
\end{equation}
and the matrix elements in $f$ are defined by
\begin{equation}\label{3dr2}
\begin{aligned}
H_{i,j}^{(n)}&=\sum_{k=0}^{i}\frac{d_{k}}{(i-k)!}(p\partial_{p}+E^{'}+n)^{i-k}\\
&\times \sum_{l=0}^{j}\frac{d_{l}^{*}}{(j-l)!}(p^{*}
\partial_{p^{*}}+E^{'*}-n)^{j-l}\frac{1}{p+p^{*}}\mid_{p=1},\\
\end{aligned}
\end{equation}
\begin{equation*}
\begin{aligned}
E^{'}_i&=\upsilon+\sqrt{\frac{6}{\alpha}}i\varphi+\sqrt{\frac{6}{\alpha}}iz-12\varsigma.
\end{aligned}
\end{equation*}

\textbf{Theorem 5.}   The 3D KP equation \eqref{3dkp1} has
 semi-rational solutions
\begin{equation}
\begin{aligned}
W=2(\ln f)_{\upsilon\upsilon},\\
\end{aligned}
\end{equation}
where
\begin{equation}\label{Bt1}
\begin{aligned}
f= \det\limits_{1\leq i,j\leq N}(H^{(n)}_{i,j})\mid_{n=0},\\
\end{aligned}
\end{equation}
and the matrix elements in $f$ are defined by
\begin{equation}\label{mm1}
\begin{aligned}
H_{i,j}^{(n)}&=(-\frac{p_i}{p^*_j})e^{E_i+E^*_j}
\sum_{k=0}^{n_i}d_{ik}(p_i\partial_{p_i}+E^{'}_i+n)^{n_i-k}\\
 \times &\sum_{l=0}^{n_j}d_{jl}^{*}(p^{*}_j
\partial_{p^{*}_j}+E^{'*}_j-n)^{n_j-l}\frac{1}{p_i+p^{*}_j}+\delta_{ij}b_{in_i}b_{jn_j}^{*},
\end{aligned}
\end{equation}
\begin{equation*}
\begin{aligned}
E_i&=p_i\upsilon+ip_i^2\sqrt{\frac{3}{2\alpha}}\varphi+ip_i^2\sqrt{\frac{3}{2\alpha}}z
-4p^3_i\varsigma+E_{i0},\\
E^{'}_i&=\upsilon+\sqrt{\frac{6}{\alpha}}i\varphi+\sqrt{\frac{6}{\alpha}}iz-12\varsigma.
\end{aligned}
\end{equation*}
Here the asterisk denotes the complex conjugation and $d_{k}$,  $d_{ik}$, $d_{jl}$,
and $d_{l}$  are arbitrary complex constants.

The proofs of Theorems 4 and
5 are similar to those of Theorems 2 and 3. It is not difficult to see
that the rational and semi-rational solutions of the 4D Fokas equation can
degenerate to the rational and semi-rational solutions of the 3D KP equation.
The corresponding transformation is as follows
\begin{equation}\label{kp-relate}
\begin{aligned}
W_{3DKPI}(\upsilon,\varphi,Z,\varsigma)&=\frac{2}{k^2_2-k^2_1}u_{4D Fokas}(x_1,x_2,y_1,y_2,t),
\end{aligned}
\end{equation}
where
$$k_1x_1+k_2x_2=\upsilon,\quad k_3y_1+k_4y_2=\sqrt{\frac{2k_3k_4}
{k_1k_2(k^2_2-k^2_1)}}(\varphi+Z),\quad
t=\frac{4}{k_2(k^2_2-k^2_1)}\varsigma.$$

\section{Conclusions} \label{7}

In this paper, the determinant expression of $N$-solitons
is constructed for the 4D Fokas equation by using the KP hierarchy
reduction method. New types of mixed solutions composed of breathers and
 V-type solitons are obtained by choosing the appropriate parameters
in Theorem 1 (see Fig. \ref{f2} and Fig. \ref{f2b}). High-order rational
solutions of the 4D Fokas equation are also derived by means of  Theorem 2,
as well as we give the condition $k_1k_2k_3k_4(k_2^2-k_1^2)>0$
to ensure that the rational solutions are smooth. We show that
the fundamental rational solution is a lump in the $(x_1,y_1)$-plane, which
is a traveling wave localized in space and time, see Fig. \ref{rational-1}. High-order rational solutions
display the interaction between several lumps in the $(x_1,y_1)$-plane, and
exhibit similar dynamical patterns to those of rogue waves in the $(1+1)$-dimensions
by altering the free parameters $c_k$ in Theorem 2
(see Fig. \ref{rational-2} and Fig. \ref{rational-3}).

Furthermore, three kinds of
new semi-rational solutions of the 4D Fokas equation are generated by introducing
differential operators $\Xi_i$ and $\mho_j$. For $N=1, n_i\geq1, \delta_{ii}=1$,
and $\delta_{ij}=0 \,(i\neq j)$ in Theorem 3, the semi-rational
solutions composed of lumps and one-soliton solutions are derived. There are
four distinct dynamical patterns of these semi-rational solutions, which are obtained by changing
the values of parameters $k_1, k_2, k_3$, and $k_4$ (see Fig. \ref{1lump-1soliton} and Fig. \ref{2lump-1soliton}).
The specific classification of these patterns is shown in Table 1. For $N\geq1,
n_i=1, \delta_{ii}=1$ and $\delta_{ij}=0 \, (i\neq j)$ in Theorem 3, the semi-rational
solutions consisting of more lumps and more solitons are also generated
(see Fig. \ref{2lump-2soliton}). Also a new kind of semi-rational
solution composed of two lumps, a breather, and a V-type soliton is
derived, for $N=2, n_i=2$, and $\delta_{ij}=1$. We point out that the interaction between the mentioned entities of
such semi-rational solution is elastic.
This kind of semi-rational solution that is illustrated in Fig. \ref{2lump-1soliton-bre}, has never been reported elsewhere,
to the best of our knowledge.

Additionally, using our rational and semi-rational solutions of
the 4D Fokas equation, we derived the rational and semi-rational solutions
of the 3D KP equation. These results indicate that the 4D
Fokas equation is a valuable multi-dimensional extension of the KP and DS equations. In addition,
this paper provides an idea for seeking the exact solutions
of high-dimensional soliton equations, and also provide a reference
for how to reduce the exact solutions of high-dimensional systems
to the exact solutions of low-dimensional ones. These results
are useful  to the study of the dynamics of nonlinear
 waves in diverse physical settings in
hydrodynamics, nonlinear optics and photonics, plasmas, quantum gases (Bose-Einstein condensates),
and solid state physics.

%%%%%%%%%%%%%%%%%%%%%%
%%%%%%%%%%%%%%%%%%%%%%%%%%%%%%%%%%%%%%%%%%%%%%%%%%%%%%%%%%%%%%%%%%%%%%%%%%%
%\textbf{Compliance with ethical standards}
\quad

\noindent \textbf{Funding}\ This work is supported by the NSF of China under Grant No. 11671219 and No. 11871446.
 \vspace{-0.5cm}\\

\section*{Compliance with ethical standards}

\noindent \textbf{Conflict of interest}\ { The authors declare that they have no conflict of interest.} \\

% BibTeX users please use one of
%\bibliographystyle{spbasic}      % basic style, author-year citations
%\bibliographystyle{spmpsci}      % mathematics and physical sciences
%\bibliographystyle{spphys}       % APS-like style for physics
%\bibliography{}   % name your BibTeX data base

% Non-BibTeX users please use

\end{document}